\documentclass[]{emulateapj}

\usepackage{graphics}

\newcommand{\mycomment}[1]{}
\newcommand{\mgii}{Mg~{\sc ii}}
\newcommand{\mgi}{Mg~{\sc i}}

\newcommand{\ovii}{{\sc O~vii}}
\newcommand{\oviii}{{\sc O~viii}}
\def\Chandra     {{\em Chandra}}

\slugcomment{Accepted for publication in Ap.J.}

\shorttitle{Chandra Observation of 3C~212}
\shortauthors{Aldcroft et al.}

\begin{document}


\title{Chandra Observation of 3C~212: a New Look at the X-ray and UV Absorbers}


\author{T. L. Aldcroft\altaffilmark{1}, 
A. Siemiginowska\altaffilmark{1}, 
M. Elvis\altaffilmark{1},
S. Mathur\altaffilmark{2},
F. Nicastro\altaffilmark{1}, and
S. S. Murray\altaffilmark{1}}
\altaffiltext{1}{Harvard-Smithsonian Center for Astrophysics, 60 Garden Street, Cambridge, MA 02138}
\altaffiltext{2}{Department of Astronomy, The Ohio State University, Columbus, OH, USA}

\email{taldcroft@cfa.harvard.edu}



\begin{abstract}
The red quasar 3C\,212 ($z=1.049$) is one of the most distant and most luminous
AGN which has shown evidence of an X-ray warm absorber.  In order to further
investigate this unusual quasar, we used \Chandra/ACIS-S to observe 3C\,212 for
19.5 ksec, resulting in a net detection of $\sim 4000$ counts.  The \Chandra\ 
data confirm the presence of an excess absorbing column $N_H \approx
4\times10^{21}$\,cm$^{-2}$ at the quasar redshift, but we find no compelling
evidence for a warm absorber.  Using both the \Chandra\  and archival ROSAT PSPC
data, we obtain very good fits for both a partially covered neutral absorber
and a low-ionization ($U = 0.03$) photo-ionized absorber.  In the ultraviolet,
3C\,212 shows a strong associated \mgii\ absorber.  Based on a moderate
resolution (80\,km\,s$^{-1}$) MMT spectrum we show that the absorber is highly
saturated and has a covering fraction less than 60\%, implying that the
absorber is truly intrinsic to the quasar.  Photo-ionization modeling of the
\mgii\ absorber yields a constraint on the ionization parameter of $U < 0.03$,
inconsistent with a warm UV/X-ray absorber.  In addition to our spectral
analysis, we find evidence in the ACIS image data for weak extended emission
surrounding the quasar as well as emission corresponding to the radio lobes at
a distance of 5\arcsec\ from 3C\,212.  The statistical significance of these
features is low, but we briefly explore the implications if the detections are
valid.

\end{abstract}


\keywords{quasars: absorption lines --- quasars:individual (3C\,212) ---
ultraviolet: galaxies --- X-rays: galaxies}


\section{Introduction}

3C\,212 (Q0855+143) is a radio-loud quasar at redshift $z = 1.049$.  It is one
of the prototype ``red quasars'' \citep{smith80} which have very faint optical
counterparts, but are bright in the near infrared (1-2$\mu$m) \citep{rieke79}.
Red quasars could be missed in typical surveys, and \citet{webster95} proposed
that up to 80\% of radio-loud quasars are hidden in the optical.  Studying the
connection between this reddening and other quasar properties can provide
insight into the structure of AGNs as well as their overall population
statistics.  If the red quasar colors are due to dust-reddening along the line
of sight \citep[see discussion in ][]{elvis94},
 perhaps associated with an obscuring torus, it is reasonable to
search for other forms of absorption.  For IR-selected red quasars from the
2MASS sample \citep{cutri01}, \citet{wilkes02} found that the implied intrinsic
absorption (based on the hardness ratio) is generally high ($N_H \sim 10^{21} -
10^{23}$\,cm$^{-2}$, but that there is no correlation between redness and
hardness ratio.  This lack of a good correlation is problematic for simple
orientation dependent models, but it may just be due to source spectral
complexity which is not modelled due to limited signal to noise.  As a
complement to broad surveys of red quasars, 3C\,212 provides the opportunity to study
a single object in detail.  In addition to being red in the optical, 3C\,212
harbors an associated \mgii\  absorption system \citep{aldcroft94} and soft X-ray
absorption detected with the ROSAT PSPC \citep{elvis94}.

The ROSAT PSPC spectrum of 3C\,212 was first published and carefully analyzed by
\citet{elvis94}.  These authors considered a wide variety of possible spectral
shapes, including absorbed and unabsorbed power-laws, pure blackbody radiation,
and a Raymond-Smith thermal plasma.  Their main conclusion regarding the X-ray
spectrum was that all models which include excess absorption were consistent with
the data.  The excellent low-energy response of ROSAT down to 0.1\,keV
positively establishes the presence of absorption, but the high energy cutoff of
just 2.4\,keV allowed degeneracy in the physical emission mechanism.

Extending the X-ray spectral analysis, \citet{mathur94} assumed a power-law as
the underlying emission and considered the possibility of a warm absorber.
They found an improvement (significant at 98\% confidence, using the F-test) in
the fit $\chi^2$ statistic when a redshifted ($z=1.049$) \ovii\  absorption edge
was added.  However, the conclusion that \citet{mathur94} reached, that
3C\,212 harbors an X-ray/UV warm absorber with ionization parameter $U \sim
0.3$, is dependent on the assumed power-law shape.

In order to further understand this interesting object, 3C\,212 was observed 
as a \Chandra\  GTO target.  In this paper we describe the results
of the observation and our spectral and imaging analysis. In 
Section~\ref{sec:observations} we describe the observations and data
reduction.  Section~\ref{sec:spec_fit} gives details of our spectral modeling
of the point source, while in Section~\ref{sec:extended} we discuss the
extended emission which may be present.  Finally, in Section~\ref{sec:mgii} we
present an optical spectrum of 3C\,212 and discuss the \mgii\  absorption, and in 
Section~\ref{sec:discussion} we present a discussion of our findings.

\mycomment{
- What is the nature of absorption for quasars at high redshift/luminosity  
  (e.g. Reeves \& Turner etc).  
- What is the luminosity of 3c212
}

\section{Observations and data reduction} 
\label{sec:observations}
\subsection{\Chandra}
3C\,212 was observed with \Chandra \citep{weisskopf02} using ACIS-S
\citep{garmire03} for 19.5 ksec on 2000-10-26 (ObsID 434).  The source was
placed on the HRMA \citep{vanspeybroeck97} optical axis at the default aimpoint
of the ACIS S3 chip.  A 1/8 sub-array readout mode with 0.5 sec integrations
was selected in order to mitigate pile-up.  Within the broadband energy range
0.3 - 10\,keV, a net total of 4006 counts in the 4\arcsec\ source region were detected,
with the X-ray celestial coordinates of the source matching the optical 3C\,212
coordinates to within $\sim 1$\arcsec.

The \Chandra\  X-ray observation data was produced by the CXCDS automatic
processing pipeline, version R4CU5UPD13.2.  In order to take advantage of
subsequent improvements in the ACIS response and gain calibration files, we
used the CIAO tool {\tt acis\_process\_events} to update the event file.  
CIAO\footnote{http://cxc.harvard.edu/ciao/} version 2.2 and
CALDB\footnote{http://cxc.harvard.edu/caldb/} version 2.10 were used in all
data analysis and processing tasks.

The spectral data reduction followed the standard CIAO 2.3
thread\footnote{http://asc.harvard.edu/ciao/threads/psextract/} to extract an
ACIS spectrum: (1) Extract source events within a 4.0\arcsec\ radius; (2)
Create the aspect histogram file; (3) Create the RMF and ARF files appropriate
to the time-dependent source position on chip.  In addition to the standard
thread, the event data were filtered on energy to use the range 0.3 - 8.0 keV,
and they were grouped to an average of 30 counts per bin.  
The actual minimum count value for each bin was chosen randomly from a gaussian
distribution having a mean of 30, standard deviation of 5, and constrained to
lie between 20 and 40.

It should be be noted that the region used to extract the spectrum includes the
possible extended emission discussed in Section~\ref{sec:extended}. However,
this component is extremely faint ($\la 30$ counts) compared to the core,
with a spectral distribution consistent with a 3\,keV thermal plasma, and
thus does not significantly affect the spectral fitting.

\begin{figure*}
\centerline{\resizebox{4.0in}{!}{\includegraphics{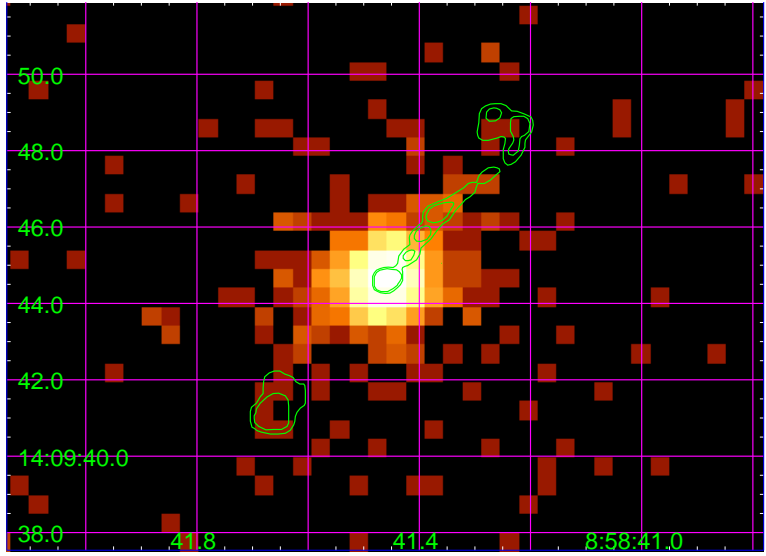}}}
\caption{18\arcsec$\times 12$\arcsec\ segment of the 20 ksec ACIS-S exposure of
3C\,212, overlayed with contours from the 3.6\,cm VLA map.  North is up and
East is to the left.}
\label{fig:image}
\end{figure*}

The \Chandra\ ACIS-S image of the source is shown in Figure~\ref{fig:image},
with 0.5 arcsec spatial binning corresponding to the ACIS CCD pixel size.  The
pixel intensities have been scaled logarithmically.  The X-ray emission does
appears circularly symmetric, as would be expected for such an on-axis
point source.  Overlayed on the image are contours from the 3.6\,cm VLA radio
map (adapted from \citep{stockton98}).  The possible extended emission and its
relation to the radio jet is discussed in Section~\ref{sec:extended}.

\subsection{ROSAT}

3C\,212 was observed with the ROSAT PSPC \citep{trumper83,pfefferman87} between
1992-05-12 to 1992-05-18 for a total of 21.6\,ksec.  Within a 2\arcmin\ source
circle a net total of 770 counts were detected.  The data were retrieved from
the ROSAT public archive and the spectrum was extracted using CIAO tools.  The
background was determined from a circular annulus with inner and outer radii of
3\arcmin\ and 6\arcmin, respectively.  We used the standard on-axis PSPC
response matrix file \texttt{pspcb\_gain2\_256.rsp}, and ignored channels 1-11 in
our analysis.  See \citet{elvis94} for further discussion of the PSPC
observation, including their Figure~1 which shows the PSPC image and the
contaminating point sources which required removal.

\section{Spectral fitting }
\label{sec:spec_fit}
Spectral fitting on the ACIS and PSPC data was carried out with CIAO 2.2
\textit{Sherpa} \citep{freeman01}, using Levenberg-Marquardt minimization and
the Gehrels $\chi^2$ statistic.  As one would expect based on current knowledge
of quasar X-ray spectra, e.g. \citet{reeves00}, the ACIS spectrum is highly
inconsistent with a pure blackbody or Raymond-Smith plasma spectral shape.  The
best fit for a redshifted blackbody with Galactic absorption ($N_H = 3.6 \times
10^{20}$~cm$^{-2}$) gives $\chi^2 = 367$ for 102 DOF, which is clearly
disallowed.  Thus the possibility of a thermal spectrum, left open with the
ROSAT data \citep{elvis94}, is no longer allowed.

In contrast, an absorbed power-law model fits the ACIS data quite well.  The
results of fitting power-law models with different absorption models are given
in Table~\ref{tab:fit}.  Using only the ACIS data, we see a significant
improvement in $\chi^2$ when comparing the model with entirely Galactic
absorption ($N_H$ free) to a model with intrinsic neutral absorption at
$z=1.049$.  To quantify the limit on absorber redshift, we fit a model with a
redshifted absorber where the redshift of the intrinsic absorber was free to
vary.  Although not well constrained, the best fit value of absorber redshift
is 1.2, which is a good match to the quasar redshift of $z=1.049$. From our fit
the 95\% confidence lower limit on the absorber redshift is $z_{abs} > 0.25$.

\begin{table*}[t]
\caption{Spectral fit parameters}
\label{tab:fit}
\small
\begin{tabular}[t]{p{1.7in}cccccc}
\hline \hline
\hfil Model \hfil  &  $\Gamma$  &  Amplitude  &  $N_{H,Gal}$ &  $N_{H,z} $ & Other & $\chi^2$ (DOF) \\
       &  (a)    & (b)                   &  (c)  &(d) & (e) \vspace{0.5ex}  \\
\hline & \vspace{-2.5ex} \\
\multicolumn{7}{c}{ACIS Only} \\
\hline  & \vspace{-2.0ex} \\
Gal $N_H$ (fixed)     & $1.27\pm0.04 $ &$2.06\pm0.07$&(0.36)& ...         & ...  & 232.7(102) \vspace{0.5ex} \\
Gal $N_H$ (free)      & $1.81\pm0.09 $ &$3.63\pm0.30$& $1.7\pm0.2$ & ...         & ...  &99.1(101) \vspace{0.5ex} \\
$N_H(z=1.049) \times$ {\sc O~vii}~edge   
                      & $1.69\pm0.09 $ &$3.11\pm0.24$&(0.36)& $3.2^{+2.2}_{-5.1}$ & $\tau_{edge} = 0.6^{+2.2}_{-3.2}$ &83.2(100) \vspace{0.5ex} \\
Part. Cov. $N_H (z=1.049)$  
                      & $1.71\pm0.08 $ &$3.17\pm0.20$&(0.36)& $4.4^{+1.3}_{-0.7}$ & $f_{PC} = 1.0^{+0.0}_{-0.1}$  & 83.3(100) \vspace{0.5ex} \\
\hline  & \vspace{-2.5ex} \\  
\multicolumn{7}{c}{ACIS + PSPC} \\
\hline  & \vspace{-2.0ex} \\  
Gal $N_H$ (fixed)     & $1.25\pm0.04 $ &$2.06\pm0.07$&(0.36)            & ...         & ...  & 271.5(125) \vspace{0.5ex} \\
Gal $N_H$ (free)      & $1.81\pm0.09 $ &$3.57\pm0.30$& $1.6\pm0.2$    & ...         & ...  & 128.0(124) \vspace{0.5ex} \\

$N_H(z=1.049)^e$        & $1.71\pm0.08$  &$3.14\pm0.20$&(0.36)            &$4.2\pm0.1$& ... &106.6(124) \vspace{0.5ex} \\
Part. Cov. $N_H (z=1.049)$  
                      & $1.72\pm0.08 $ &$3.19\pm0.22$&(0.36)            &$4.8^{+1.5}_{-1.2}$ & $f_{PC}=0.94^{+0.06}_{-0.08}$ & 105.7(123) \vspace{0.5ex} \\
$N_H(z=1.049) \times$ {\sc O~vii}~edge    
                      & $1.69\pm0.08$  &$3.07\pm0.18$&(0.36)            &$2.3^{+2.1}_{-0.8}$ & $\tau_{edge}=1.3^{+0.6}_{-1.1}$ &102.3(123) \vspace{0.5ex} \\
\hline  & \vspace{-2.5ex} \\  
\multicolumn{7}{c}{ACIS + PSPC ($\Gamma_{\rm PSPC} = \Gamma_{\rm ACIS} + 0.4$)} \\
\hline  & \vspace{-2.0ex} \\  
Gal $N_H$ (fixed)     & $1.22\pm0.04 $ &$2.03\pm0.07$&(0.36)            & ...         & ...  & 316.0(125) \vspace{0.5ex} \\

Part. Cov. $N_H (z=1.049)$  
                      & $1.72\pm0.08 $ &$3.20\pm0.21$&(0.36)            &$4.8^{+1.0}_{-1.2}$ & $f_{PC}=0.96^{+0.06}_{-0.07}$ & 96.5(123) \vspace{0.5ex} \\
$N_H(z=1.049) \times$ {\sc O~vii}~edge    
                      & $1.70\pm0.08$  &$3.14\pm0.18$&(0.36)            &$3.5^{+2.0}_{-1.1}$ & $\tau_{edge}=0.7^{+0.7}_{-1.3}$ &95.6(123) \vspace{0.5ex} \\
Warm Abs. ($z=1.049$) & $1.71\pm0.06$  &$3.20\pm0.23$&(0.36)            &$5.0^{+1.3}_{-1.0}$ & $\log U \le -0.8$  &99.0(122) \vspace{0.5ex} \\

\hline \vspace{-1.5ex} \\ 
\multicolumn{7}{c}{
\parbox{6.5in}{
{\sc Notes:}  Uncertainties are 90\% 
confidence limits.  Fit values in parentheses are frozen. 
(a) Power law photon index.  (b) Power law normalization in units of $10^{-4}$\,photons\,cm$^{-2}$\,s$^{-1}$\,keV$^{-1}$ at 1\,keV
(c) Absorbing column ($10^{21}$\,cm$^{-2}$) at $z=0$. (d) Absorbing column ($10^{21}$\,cm$^{-2}$) at quasar redshift.  
(e) $f_{PC} = $ partial covering fraction, $\tau_{edge} = $ edge optical depth, $U = $ CLOUDY ionization parameter.
(f) Best fit parameter values found by \citet{mathur94}.
}}
\end{tabular}
\end{table*}

Unfortunately, the response and calibration accuracy of ACIS-S fall off rapidly
below about 0.5\,keV, making it difficult to distinguish between different
absorption scenarios.  The key diagnostic edges of \ovii\ and \oviii\ 
are shifted to 0.35~keV and 0.41~keV respectively at z=1.049.  The low energy
response of the PSPC data can help.  Since the observations were taken about 8
years apart, variability is an obvious concern.  However, we find only a
moderate increase of 37\% in source flux (0.5 - 2.5~keV) from the ROSAT to
\Chandra\  observation, and the power-law spectral index for non-BL Lac type
quasars does not typically show large variability \citep{sambruna97}.

\begin{figure*}
\epsscale{2.0}
\resizebox{3.2in}{!}{\includegraphics{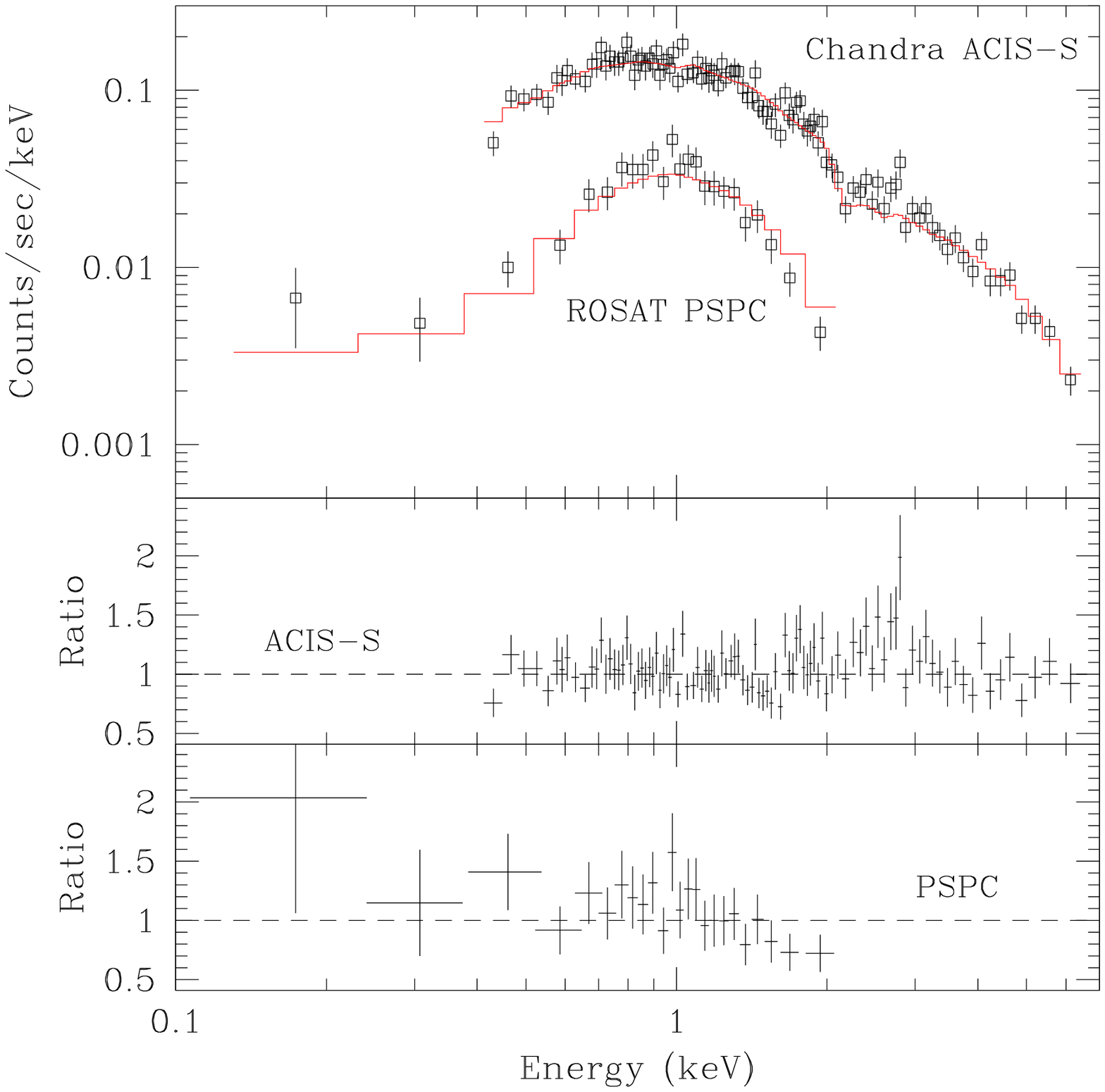}}
\resizebox{3.2in}{!}{\includegraphics{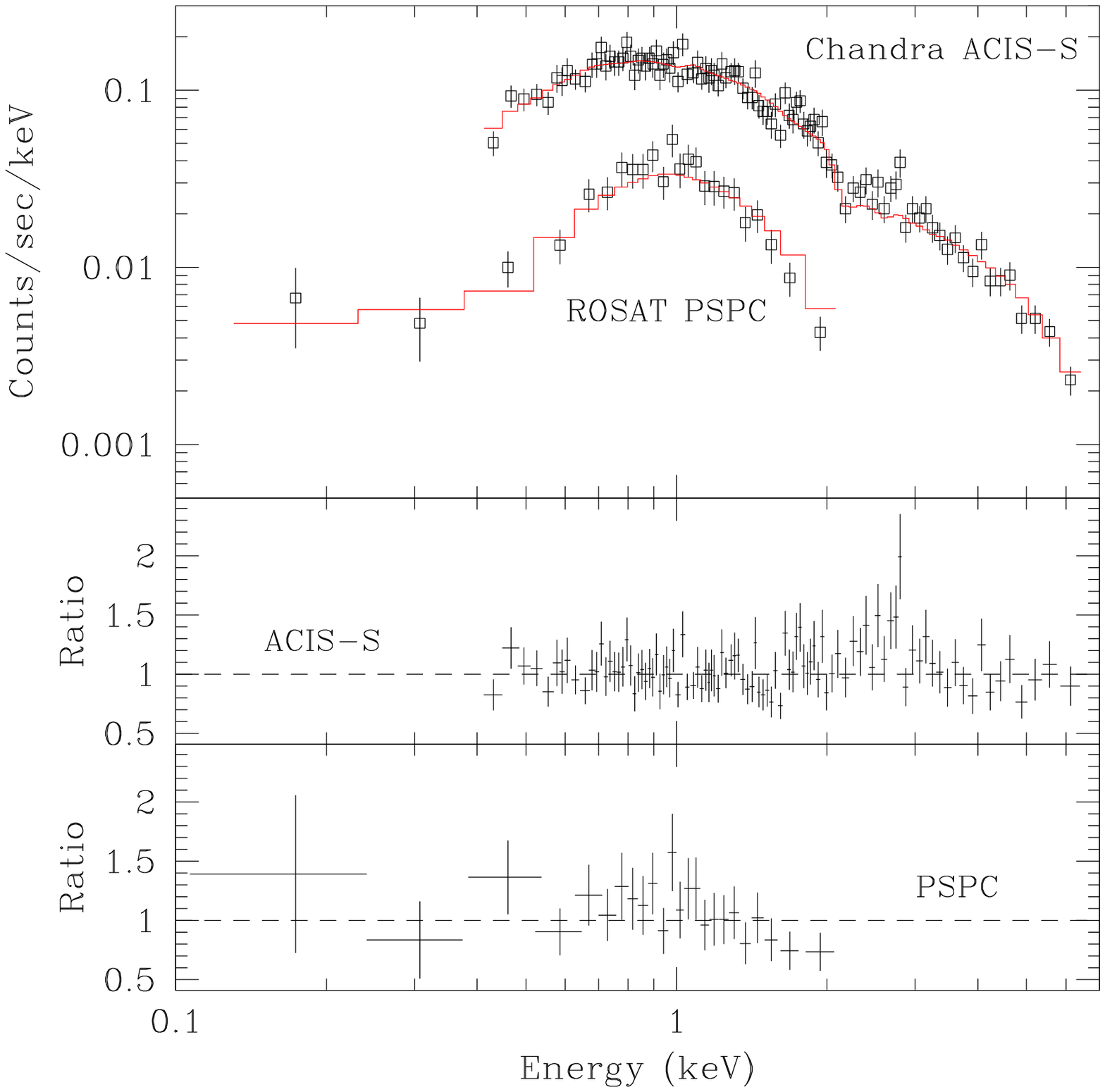}}
\caption{Results of simultaneous spectral fitting of ACIS and PSPC data using
two models with intrinsic ($z=1.049$) absorption.  The left plot shows the fit
for a partially covered absorber model ($f_{PC}=0.94^{+0.06}_{-0.08}$) 
and the right plot shows the fit for an
absorber with an \ovii\ edge. The lower panels in each
of the two plots show the ratio of data/model.}
\label{fig:spectrum}
\label{fig:wabs_redshift}
\epsscale{1.0}
\end{figure*}

The spectral fits and
residuals for the remaining four models, which have absorption at $z=1.049$,
are plotted in Figure~\ref{fig:spectrum}.

The results of simultaneous fitting of the ACIS + PSPC data sets are shown in
the second section of Table~\ref{tab:fit}.  Here we fit exactly the same model
to each data set, but add a relative normalization parameter which is free to
vary.  The first two models, which include only absorption at $z=0$, give much
larger values of $\chi^2$ than the intrinsic absorption models, and can
be excluded.  A simple power-law with a neutral absorbing column of $4.2\times
10^{21}\,$cm$^{-2}$ at the quasar redshift gives a good fit to the data, giving
a $\chi^2$ of 106.6 for 124 DOF.  Adding a partial-covering factor
(Figure~\ref{fig:spectrum}a) improves the $\chi^2$ only marginally, although
the very lowest energy PSPC points are fit better.  A larger improvement in the
fit is found by adding an absorbing \ovii\ edge at the quasar redshift
(Figure~\ref{fig:spectrum}b).  This confirms the finding of \citet{mathur94},
though with the addition of the ACIS data and a well-constrained power-law
photon index, we now find a smaller optical depth (1.3 versus 3.9) with a
somewhat improved detection confidence of 2-$\sigma$.  These values can be
compared with the best fit values found by \citet{mathur94} of $\Gamma = 2.5$,
$N_{H,z} = 3.6 \times 10^{21}$, and $\tau_{\rm OVII } = 3.9$.  The power-law derived
using only PSPC data is too steep to be compatible with the ACIS data.

\begin{figure}
\centerline{
\resizebox{3.5in}{!}{\includegraphics{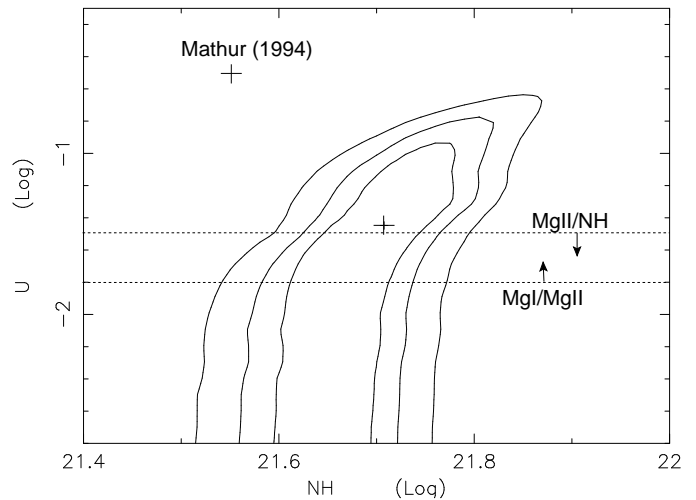}}}
\caption{Plot of 1,2,3-$\sigma$ contours for a warm
absorber model fit of 3C\,212 data, indicating a low-ionization or neutral
intrinsic absorber.  The dashed lines show constraints on the ionization
parameter assuming a one-zone UV/X-ray absorber.  The limits are based on
the \mgi/\mgii\ ratio (lower) and the \mgii/$N_{H}$(X-ray) ratio (upper).
We also show the warm absorber parameters found by \citet{mathur94}.}
\label{fig:warm_abs}
\end{figure}

The third section of Table~\ref{tab:fit} shows the results of simultaneous
fitting of the ACIS and PSPC data sets, where we have imposed the constraint on
spectral indices of $\Gamma_{\rm PSPC} = \Gamma_{\rm ACIS} + 0.4$.  This is
motivated by results which indicate that AGN spectral indices derived with the
ROSAT PSPC are systematically steeper ($\Delta \Gamma \approx 0.4$) than those
from the ASCA SIS and the Einstein IPC \citep{fiore94,laor94,iwasawa99}.
This applies for AGN data fit over the same energy band (0.5 - 2~keV) using a
variety of spectal models, suggesting a problem in the effective area curve
\citep{iwasawa99}.  For the models with intrinsic absorption, we find a
significant improvement in the fit $\chi^2$, but only small changes in the fit
parameters (compared to the results for $\Gamma_{\rm PSPC} = \Gamma_{\rm
ACIS}$).  The most notable change is that the best \ovii\ edge opacity is
decreased to $\tau_{edge}=0.7^{+0.7}_{-1.3}$, implying that the edge is only
marginally detected.  We also fit the data with a warm absorber model which is
based on CLOUDY \citep[Version 94;][]{ferland96} calculations.  The CLOUDY warm
absorber model assumes a photo-ionized plasma with Solar metallicity, density
of $10^{9}$~cm$^{-3}$, and a standard AGN ionizing continuum \citet{mathews87}.
A contour plot showing 1, 2, and $3-\sigma$ confidence intervals for absorber
$N_H$ versus ionization parameter $U$ is given in Figure~\ref{fig:warm_abs}.
This points to a neutral or low-ionization absorber, and the ionization
parameter of $\log U = -0.5$ found by \citet{mathur94} is just outside our
3-$\sigma$ contour interval.

One feature which is evident in the ACIS ratio plots in
Figure~\ref{fig:spectrum} is a significant bump at $\sim 2.5$\,keV with a width
of around 1\,keV.  This bump is consistent with the very broad iron
line emission reported in MCG-6-30-15 during one 26\,ksec interval
\citep{iwasawa96}, which \citet{reynolds97} interpreted as evidence for
a relativistic accretion disk surrounding a maximally rotating Kerr black hole.
However, we have noticed similar residuals at this level ($\sim 30$\%) at the
same observed frame energy are seen in other \Chandra\ quasar spectra, and we
conclude that the excess is most likely instrumental in nature.

\section{Extended emission }
\label{sec:extended}
Figure~\ref{fig:image} shows that the 3C\,212 X-ray emission appears slightly
elongated along the axis defined by the radio jet.  Furthermore, there is
evidence of X-ray emission which is correlated with the hot spots of radio
emission, both in the jet and at the lobes.

\begin{figure}
\centerline{\resizebox{3.5in}{!}{\includegraphics{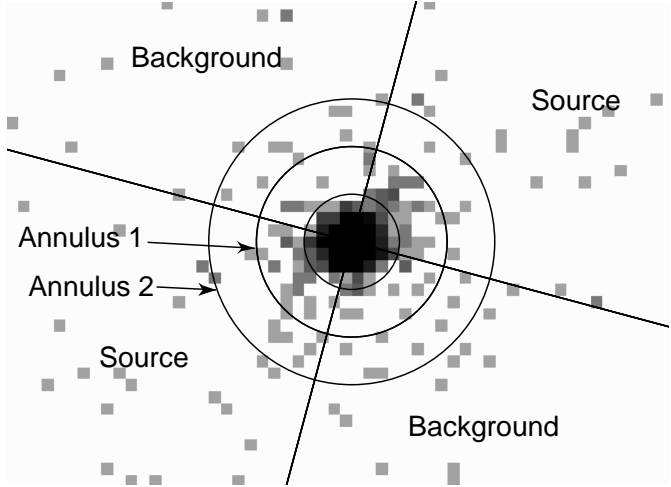}}}
\caption{Soft-band image (0.3 - 2.5 keV) of 3C\,212, showing apparent
extended emission in the upper-right and lower-left quadrants.  Net extended
counts for each of two annuli were calculated using the indicated source and
background wedges.  The inner annulus (2-4\arcsec) has a net of
$30\pm10$\,counts, while the outer annulus (4-6\arcsec) has $-2\pm7$\,counts.
In the figure North is up and East is to the left.}
\label{fig:extended}
\end{figure}

Looking first at the elongation, Figure~\ref{fig:extended} shows that the X-ray
image is slightly elongated in the soft band (0.3 - 2.5 keV).  This is
inconsistent with the expectation for a point source, given that the HRMA PSF
is essentially circular at the location of this source.  In the hard band (2.5
- 8.0 keV) no such feature is present, but it should be noted that even if the
source were truly extended in both hard and soft band, the elongation would not
be detectable the hard band due to the smaller number of counts.

To estimate the formal significance and luminosity of the possible elongated
extended emission, we use a simple method of dividing the image into quadrants
which correspond to extended source counts and background counts
\citep{elvis83}.  The geometry and regions are illustrated is
Figure~\ref{fig:extended}.  We exclude the bulk of the central point source
counts within 2\arcsec, and only examine two annular rings at radii of
2-4\arcsec\ and 4-6\arcsec.  The net counts are calculated simply by summing
the counts in the two source wedges and subtracting the counts from the
background wedges.  For the inner annulus we find a net of $30\pm10$\,counts,
while the outer annulus is consistent with no extended flux ($-2\pm7$\,counts).
As a consistency check of the method, we carried out the same analysis for a
\Chandra\  ACIS-S observation of AO\,0235+164 (ObsId 884). Fortuitously, this
source is very similar to 3C\,212 in spectral shape, observed flux, readout
mode, and aimpoint on the CCD, which makes possible a direct comparison of the
spatial flux distribution for the two observations.  In AO\,0235+164, we find
that both annular rings are consistent with no asymmetric extended flux as seen
in 3C\,212.

Another possible origin for the extended flux would be the ACIS
readout streak associated with the clocking of the ACIS CCDs during
readout.  This must be considered because the direction of the
extension corresponds with the readout direction.  However, careful
examination of the image reveals no evidence of a readout streak.
Furthermore, the observation of AO\,0235+164, which is a close
analog to 3C\,212 in both count rate (within 10\%) and CCD frame time
(0.5\,sec), shows no elongation.

If the excess of 30 photons in 3C\,212 originates in a thermal bremstraulung
plasma at $kT = 1.5$~keV, the flux in the 0.3 - 2.5~keV band (observed frame)
is $f = 6\times 10^{-15}$~erg\,s$^{-1}$\,cm$^{-2}$.  The same value of flux is
also found in the 0.3 - 2.5~keV band in the quasar rest frame. At the redshift
of 3C\,212, this flux corresponds to a luminosity of $L = 3.6\times
10^{43}$~erg\,s$^{-1}$ (for $H_0 = 71$~km\,s$^{-1}$\,Mpc$^{-1}$,
$\Omega_{\Lambda} = 0.73$, and $\Omega_{M} = 0.27$).

\begin{figure}
\epsscale{1.0}
\centerline{\resizebox{3.5in}{!}{\rotatebox{90}{\includegraphics{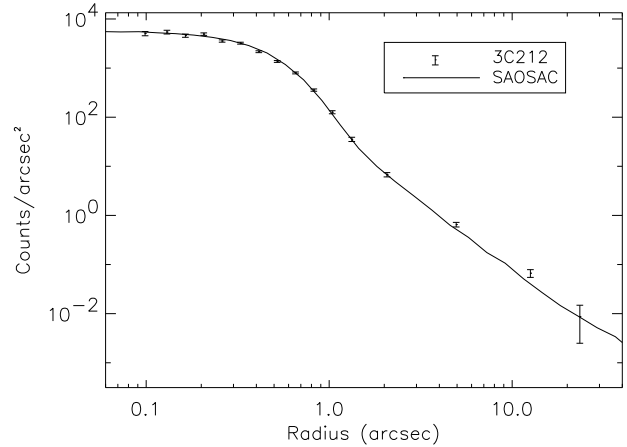}}}}
\caption{Radial profile of
of 3C\,212 (points with vertical error bars) and the SAOSAC model profile (line).
There is no evidence for significant extended emission.
}
\label{fig:profile}
\end{figure}

A different way of searching for extended emission is to compare the
observed radial profile to the predicted profile generated with the
detailed SAOSAC HRMA raytrace model developed by the CXC calibration
group \citep{tibbets00}.  An input spectrum corresponding to the
partially-covered neutral absorber model was supplied to SAOSAC, which
simulated a large number of rays being imaged through the HRMA.  The
resultant set of rays were then processed through a simple ACIS
simulator which accounts for spacecraft dither and aspect blurring,
ACIS pixelation, and the ACIS quantum efficiency.  The resulting
radial profile is shown as the solid line in Figure~\ref{fig:profile},
with the corresponding 3C\,212 points shown as vertical error bars.
By this method we also see a slight excess at radii of 3 -
10\arcsec, although this is within the systematic uncertainty of the
SAOSAC model (D. Jerius, personal communication).  Nevertheless, the
profile clearly demonstrates that 3C\,212 is not embedded in an X-ray
luminous cluster.

\subsection{Jet emission?}
The contours in Figure~\ref{fig:image} correspond to the 3.6\,cm VLA image,
adapted from Figure~1 in \citet{stockton98}.  Within the contours of the
Southern and Northern lobes, we see 5 and 3 photons respectively.  We can
estimate the significance of these numbers by looking at the background rate
for an annulus centered on the quasar at the radius of the lobes ($\sim
5$\arcsec).  For the Southern lobe, the expected counts within a 1.9\arcsec\ 
diameter circle is 1.4, which gives a probability of 1.6\% of seeing 5 or more
counts.  For the Northern lobe, we expect 0.9 counts in a 1.5\arcsec\  diameter
circle, so the chance probability of seeing 3 or more counts is 6.3\%.
Considering both lobes at once, the probability of seeing 8 or more counts is
0.26\%.  Since these locations are pre-defined by the radio lobes these are not
{\em a posteriori} statistics and do suggest a detection.

These probability estimates must be taken with a grain of salt, and certainly the
hot spot X-ray fluxes are highly uncertain, but it is nevertheless worth
calculating the basic jet characteristics assuming the counts are truly due to
X-ray jet emission.  In this case each component would have a luminosity on the
order of $L \sim 10^{42}$~erg\,s$^{-1}$.  The flux density at 5~GHz for the
Southern and Northern hot spots are 13~mJy and 74~mJy respectively
\citep{akujor91}. The slope of the spectrum between the radio and X-ray
emission (F$_{South}$(1~keV)$ = 1.93\times 10^{-15}$~ergs~cm$^{-2}$~sec$^{-1}$
and F$_{North}$(1~keV)$ = 1.17\times10^{-15}$~ergs~cm$^{-2}$~sec$^{-1}$) gives
spectral indices of $\alpha _{South} = 1.04$ and $\alpha _{North} = 0.94$
respectively ($F\sim
\nu ^{-\alpha}$).  Note that the long lever arm from X-ray to radio makes
$\alpha$ only weakly dependent on the precise X-ray flux. Hot spots usually
emit X-rays through synchrotron or synchrotron self-Compton process
\citep[SSC, jet synchrotron photons are inverse Compton scattered of the
relativistic particles in the jet, see recent reviews by
][]{harris02,wilson03}. Based on the ratio of the radio to X-ray flux densities
both processes can account for the observed X-rays from the hot spots in
3C\,212.  These hot spots are at high redshift ($z=1.049$) and if the bulk motion
of the gas in the hot spot is high (Lorentz factors of a few) the X-ray
emission may be due to the scattering of CMB photons on the relativistic
electrons.  This process is more effective at high redshift since the photon
energy density of the CMB scales as $(1+z)^4$
\citep[e.g.][]{hardcastle02,siemiginowska02}. 

\section{Associated \mgii\  absorber}
\label{sec:mgii}
\begin{figure*}
\centerline{\resizebox{7.0in}{!}{
\includegraphics{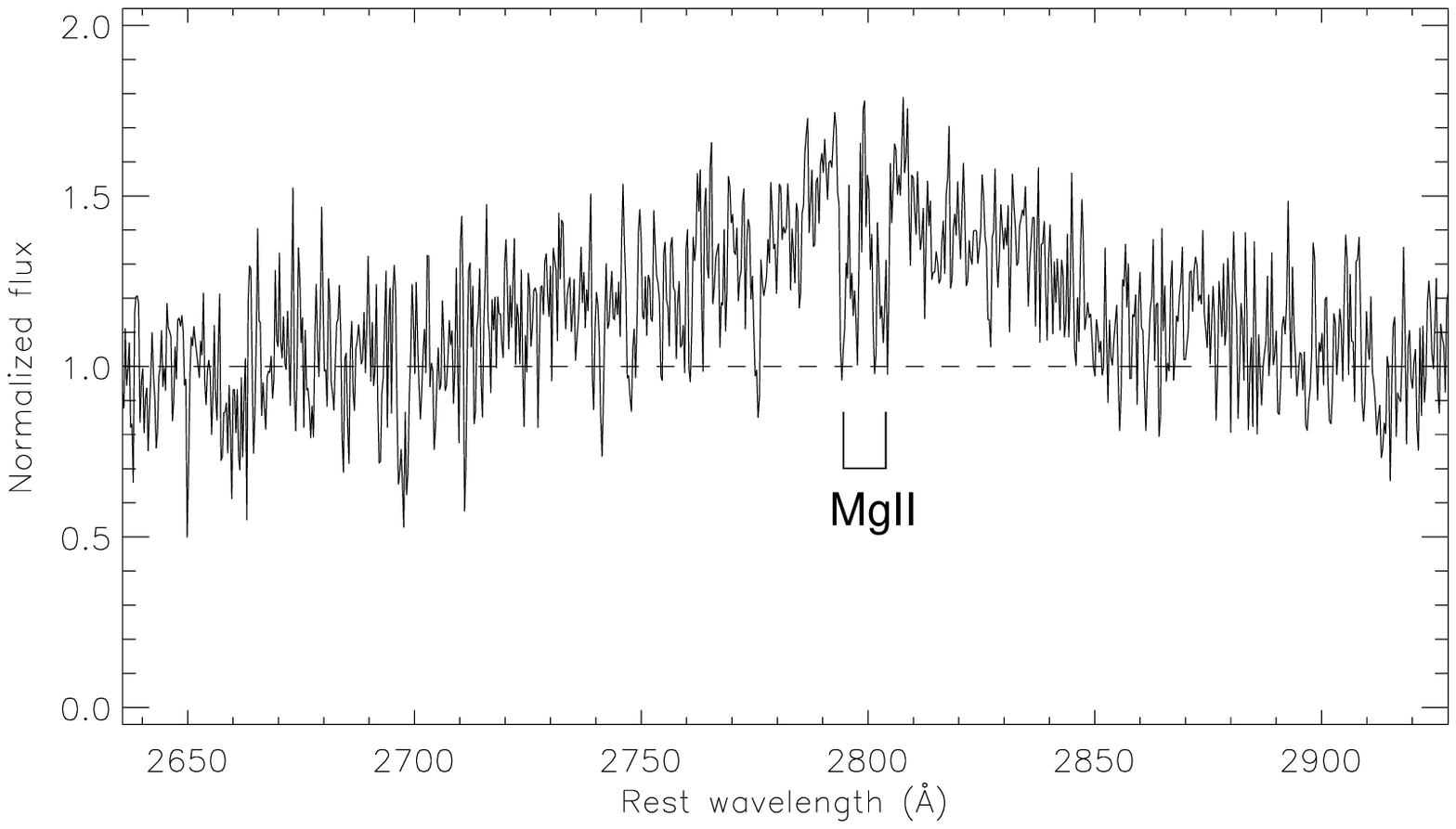}
\includegraphics{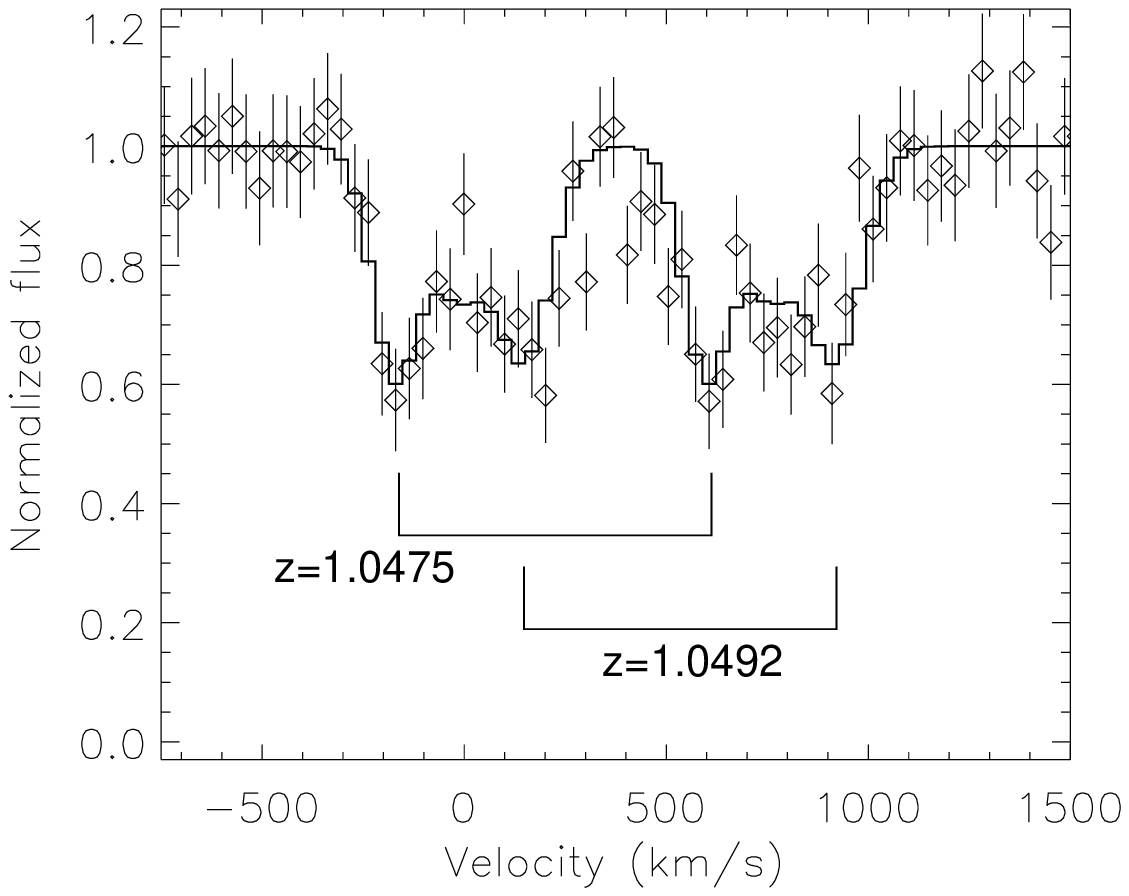}}}
\caption{Optical spectrum of 3C\,212 near the \mgii\  emission line.  Left panel
shows the spectrum normalized by the underlying power-law continuum (dashed
line).  Right panel shows a close-up of the associated \mgii\  absorption complex,
normalized by the continuum + BELR flux.  The velocity scale applies to the
\mgii(2796) line and is relative to the estimated quasar redshift $z=1.0486$
derived from [NeV](3426).  The solid line is indicates a fit to the profile
using three gaussians.  }
\label{fig:mgii_spec}
\label{fig:mgii_fit}
\end{figure*}

3C\,212 is known to harbor an associated \mgii\  absorption complex, with
components at $z=1.0491$ and $z=1.0477$ \citep{aldcroft94}.
Figure~\ref{fig:mgii_spec} shows the \mgii\  emission region from a previously
unpublished spectrum (resolution 85\,km\,s$^{-1}$) taken with the
Multiple Mirror Telescope red channel echellette on 28-Dec-94.  The spectrum
was reduced and flux calibrated in the standard manner using the IRAF {\tt
noao.imred.echelle} package.  For reference we also show a straight-line fit to
the underlying quasar power-law continuum.  Given that the \mgii\  absorption
complex does not extend below the continuum level, these data do not require
that the absorbing material obscures the continuum source.  However, a geometry
which covers 100\% of the broad emission line (BEL) region without covering the
small central continuum seems highly unlikely.

In the right panel of Figure~\ref{fig:mgii_fit} we show just the \mgii\
absorption complex, normalized by the continuum + BELR flux.  The solid line
shows a fit to the absorption profile using three gaussian \mgii\ absorbers,
where we have fixed the \mgii\ components (at 2796\,\AA\ and 2803\,\AA) to have
exactly the same amplitude and FWHM. The doublet line oscillator strengths have
a 2:1 ratio, so the 1:1 ratio seen in the line optical depths over the profile
demonstrates that the spectrum is consistent with fully saturated line
components.  Thus the profile essentially represents variation in covering
fraction with outflow velocity.  Assuming an optical depth $\tau \ge 3$ over
the width of the profile implies a lower limit on the \mgii\ column density of
$N({\rm MgII}) \ga 1.1\times 10^{15}$\,cm$^{-2}$.  If both the continuum and
BELR are absorbed, then the covering fraction is $C_f \la 0.4$; if only the
continuum is absorbed then we have $C_f \la 0.6$.  One immediate conclusion is
that the \mgii\ absorber in 3C\,212 is intrinsic to the quasar: partial
covering implies that the absorber size is comparable to the quasar UV emitting
region and hence much less than the size of typical intervening galactic ISM
clouds.

Complementary data supporting this conclusion is found in \citet{stockton98},
who used Keck LRIS to carry out a comprehensive redshift survey of faint
galaxies in the 3C\,212 field.  They found only one galaxy within 100\arcsec\
with a redshift very near that of 3C\,212, and that galaxy redshift of
$z=1.053$ is inconsistent with the \mgii\ absorption redshift range of
$z=1.0475$ to $z=1.0492$.

The observation of a saturated \mgii\ complex and non-detection of the
corresponding \mgi\ line allows us to constrain the ionization level of the
absorbing cloud material.  We used CLOUDY \citep[Version 94;][]{ferland96} to
calculate the Mg ionization fractions as a function of the dimensionless
incident ionization parameter $U$.  In all models we used a particle density of
$10^{7}$~cm$^{-3}$, a total absorbing column of $3.5\times10^{21}$~cm$^{-2}$
(from the \ovii\ edge model in Table~\ref{tab:fit}),
and solar abundances.  In order to investigate the effect of the shape of the
ionizing continuum, we used three different input spectral energy distributions
(SED): (1) the standard CLOUDY radio-quiet AGN table model based on
\citet{mathews87}; (2) the average SED for radio-loud quasars \citep{elvis94b};
(3) the SED for the radio-loud quasar 3C351 \citep{elvis94}.  We find that for
the low-ionization \mgi\ and \mgii\ transitions, the results for the three
different SEDs are nearly identical. The non-detection of
the \mgi\ line at 2852~\AA\ (rest) requires that $N({\rm MgI}) \la 1.6\times
10^{12}$\,cm$^{-2}$ (at 2-$\sigma$ confidence) and hence $\log U > -1.8$.

If we assume that the X-ray and UV absorption are due to the same material
(i.e. a one-zone model), then the observed lower limit on the \mgii\ column of
$N({\rm MgII}) \ga 1.1\times 10^{15}$\,cm$^{-2}$ translates into an upper limit
on the ionization parameter $\log U < -1.5$.  
These constraints conflict with \citet{mathur94}, who argued that the observed
\mgii\ is consistent with a moderately high ionization level of $\log U =
-0.5$.  Part of the discrepancy is the assumed \mgii\ column, since they
considered a range from $\approx 2\times 10^{13}$ to $2\times
10^{14}$~cm$^{-2}$ (based on lower quality optical spectroscopy).  However,
even at this extreme low column, our models indicate that $\log U < -0.8$.

An important point to note from the \mgii\ profile in Figure~\ref{fig:mgii_fit}
is that although part of the absorbing material appears to be inflowing at a
velocity of up to $\sim 200$\,km\,s$^{-1}$, this may not be correct. The
assumed quasar redshift of $z=1.0486$ was derived from the [NeV](3426) line.
However, in Seyfert galaxies this forbidden high-ionization line is found to be
blueshifted relative to the lower ionization lines by as much as
450~km\,s$^{-1}$ \citep{erkens97}.  This effect is correlated with the
forbidden line FWHM.  Given the rather broad profile of [NeV] in 3C\,212 (over
700~km\,s$^{-1}$), the \mgii\ absorbing
material could well be outflowing.

\section{Discussion}
\label{sec:discussion}

X-ray grating spectra from \Chandra\  and XMM-Newton have confirmed that ionized
X-ray ``warm'' absorbers share an outflow velocity with the UV absorbers seen
in the same AGN \citep[e.g.][]{kaastra02,kaspi02,collinge01}, although the
X-ray spectra lack the resolution to distinguish all the components seen in the
STIS and FUSE UV spectra.  These X-ray spectra are consistent with the
high-ionization wind picture deduced by
\citet{mathur95}.  However, in 3C\,212 the UV absorber shows 
low-ionization \mgii\ with $N$(\mgii)$ \ga 10^{15}$\,cm$^{-2}$ 
coexisting with an X-ray absorber with $N_H \sim 4\times
10^{21}$\,cm$^{-2}$ and ionization $U < 0.16$.  The two absorbers can only be
from the same gas over the relatively narrow range of ionization parameter
$0.016 < U < 0.03$.  

However, the available data do not require such a single-zone solution.
\citet{elvis00} proposed a two-phase medium in the wind with the cooler phase
corresponding to the broad emission line region gas.  The ionization of this
gas is typically $\log U \sim -1$ to $-0.7$, well in line with our observed
limit (Figure~\ref{fig:warm_abs}).  \citet{krolik01} instead
suggest a multi-phase medium spanning a range of 30 in temperature.  X-ray
grating spectra do indicate at least two different ionization states in some
AGN winds \citep{kaastra02}.  The large variation in ionization within single
objects inferred from high-resolution UV spectroscopy may also imply multiple
absorbers and stratified ionization \citep[e.g. NGC5548][]{crenshaw99},
although some of the apparent complexity seems to be analysis dependent
\citep{krongold03}.  In the particular case of FIRST J1044+3656, using a
multi-phase model allows a much more physically plausible interpretation of the
excited state absorption data, moving the absorber distance from 700~pc
\citep[single-zone,][]{dekool01} to 4~pc \citep{everett02}.

A two-phase absorber is likely in 3C\,212, since the low-ionization (\mgii)
material only partially covers ($C_f \la 0.6$) the continuum source, while the
X-ray absorber appears to cover the X-ray source completely ($C_f > 0.89$ at
90\% confidence).  If this is the case, then the upper and lower-limit
constraints (Figure~\ref{fig:warm_abs}) on the ionization of the UV absorbing
material do not apply.  The upper limit depends on the ratio of \mgii/$N_H$, so
decoupling the absorbing regions clearly invalidates our calculation.  For the
lower limit, as one moves to lower ionization parameter the relative population
abundance of \mgii\ increases faster than that of \mgi. By considering a lower
UV absorbing cloud column (without the constraint that it matches the total
X-ray absorbing column), the ionization can be much lower while maintaining the
same observed \mgi\ and \mgii\ equivalent widths. This is analogous to typical
intervening \mgii\ absorbers which arise in the ISM of foreground galaxies
and typically do not show \mgi\ absorption \citep{steidel92}.
\citet{mathur94} point out that the lack of detected 21\,cm absorption at the
quasar redshift rules out a mostly-neutral absorber, but this assumes both a
one-zone system as well as a 21\,cm spin temperature of 100\,K.  The latter
assumption is particularly unlikely in the vicinity of the quasar central black
hole.

If instead both UV and X-ray absorbers have low ionization then we are probably
seeing a different type of absorber, one which is commonly found in absorbed
type 1.8 and 1.9 AGN \citep{risaliti01}.  These AGN also have reddened
continua, and broad emission lines with their broad components of H~$\beta$
(Seyfert 1.8's) or H~$\alpha$ and H~$\beta$ (Seyfert 1.9's) being suppressed.
This strengthens the analogy with the red quasar 3C\,212.  These low-ionization
absorbers may be unrelated to the high-ionization winds.
%

The uncertain \mgii\ velocity with respect to the nucleus leaves in doubt
whether the low-ionization absorber is at rest (as in a torus) or in outflow as
in the high-ionization winds.  High-velocity low-ionization winds are seen in 
LoBALs (Low Ionization Broad Absorption Line systems).
\citet{risaliti02} find that the $N_H({\rm eq})$ of the low-ionization
absorbers in Seyfert 1.8, 1.9's varies in virtually all cases, and does so on
timescales of a few months.  They suggest that a wind similar to the
high-ionization absorbers but extending to larger radii, and so lower
ionization, could explain these absorbers as well.

High resolution X-ray spectra of 3C\,212 down to energies of $\sim 0.1$~keV
could discriminate bewteen a single and a two-phase medium, measure the
absorber covering factor, and might measure the velocity of the absorbing
material.  This should be feasible with the next generation of large area X-ray
telescopes.

\acknowledgments

The authors would like thank Susan Ridgway for helpful discussions and for
providing a VLA map of 3C\,212.  We would also like to thank the entire \Chandra\ 
team for making possible the observations and data analysis. This work was
supported by NASA grant NAS8-39073.



\clearpage


\clearpage

\end{document}